\documentstyle[sprocl,epsf]{article}
\epsfverbosetrue

\def\Journal#1#2#3#4{{#1}{\bf #2}, #3 (#4)}

\def\NPB{{\em Nucl.~Phys.~}B}
\def\PLB{{\em Phys.~Lett.~}B}
\def\PRL{\em Phys.~Rev.~Lett.~}
\def\PRD{{\em Phys.~Rev.~}D}
\def\ZPC{{\em Z.~Phys.~}C}

\def\EPJ{{\em Eur.~Phys.~J.~}C}

\def\be{\begin{equation}}
\def\ee{\end{equation}}
\def\bea{\begin{eqnarray}}
\def\eea{\end{eqnarray}}

\begin{document}

\title{THINKING ABOUT TOP \\ within the Standard Model~\footnote{Presented
at the Thinkshop on Top-Quark Physics for the Tevatron Run II.}} 

\author{S.~WILLENBROCK}

\address{Department of Physics, University of Illinois at Urbana-Champaign, \\
1110 West Green Street, Urbana, IL 61801}


\maketitle\abstracts{I present an overview of standard-model
top-quark physics at the Fermilab Tevatron.
Topics discussed include the top-quark mass, weak interaction, 
strong interaction, and rare decays.}

\section{Why think about top?}

Before we begin thinking about the top quark, let's clarify why we should 
think about it.  The main reason is that the near future holds the promise of a
large number of top quarks at the Fermilab Tevatron.\footnote{In this talk
I restrict my attention to top-quark physics at the Tevatron.  I do not
consider top-quark physics at the LHC or future lepton colliders.}
For example, let's consider some of the cleanest
top-quark events, $t\bar t \to W + 4j$, with at least one $b$ 
tag.\footnote{The $W$ is identified via its leptonic decay.}  These events are 
fully reconstructable and have very little background.  In the Run I data
($\sqrt S = 1.8$ TeV, 100 pb$^{-1}$), each experiment had about 25 such 
events.\cite{MTOP}\footnote{For example, CDF had 34 such events, 
of which about 8 are thought to be background.}
There are expected to be about 1000 events per experiment in Run II 
($\sqrt S = 2$ TeV, 2 fb$^{-1}$), due mostly to the factor of 20 increase in 
integrated luminosity, but also due to the 37\% increase in production 
cross section at 
$\sqrt S = 2$ TeV and the increased acceptance for top-quark events.  
Further running beyond
Run II could deliver as much as 30 fb$^{-1}$ (``Run III''), which corresponds
to about 15,000 events.  The large number of events produced in Runs II and III
will allow a detailed scrutiny of the properties of the 
top quark.

What are the chances that a close inspection of the properties of the 
top quark will yield surprises?  One way to address this question is to 
consider the top-quark's SU(2) partner, the $b$ quark.  The $b$ quark
was discovered in 1977,\cite{BDISC} and in 1983 it yielded its first surprise:
its lifetime was found to be much longer than expected.\cite{BLIFE}  This is a 
consequence of the fact that $V_{cb} \ll V_{us}$, something which could not
have been anticipated.  The large $b$ lifetime has a number of very desirable 
consequences, such as large $B^0$-$\bar {B}^0$ mixing, 
large $CP$ violation, enhanced rare decays, and the ability to tag $b$ jets 
via a secondary vertex using a silicon vertex detector.

The top quark was discovered in 1995,\cite{TOP} and has already yielded its
first surprise: the large value of its mass, approximately 174 GeV.  
Fifteen years ago, there were few
who would have guessed its mass would be so large.  A detailed scrutiny of
the top-quark's properties will reveal whether there are more surprises 
in top-quark physics.

\begin{figure}[t]
\begin{center}
\setlength{\unitlength}{0.240900pt}
\ifx\plotpoint\undefined\newsavebox{\plotpoint}\fi
\sbox{\plotpoint}{\rule[-0.200pt]{0.400pt}{0.400pt}}%
\begin{picture}(974,809)(0,0)
\font\gnuplot=cmr10 at 10pt
\gnuplot
\sbox{\plotpoint}{\rule[-0.200pt]{0.400pt}{0.400pt}}%
\put(221.0,103.0){\rule[-0.200pt]{4.818pt}{0.400pt}}
\put(199,103){\makebox(0,0)[r]{0.001}}
\put(890.0,103.0){\rule[-0.200pt]{4.818pt}{0.400pt}}
\put(221.0,219.0){\rule[-0.200pt]{4.818pt}{0.400pt}}
\put(199,219){\makebox(0,0)[r]{0.01}}
\put(890.0,219.0){\rule[-0.200pt]{4.818pt}{0.400pt}}
\put(221.0,335.0){\rule[-0.200pt]{4.818pt}{0.400pt}}
\put(199,335){\makebox(0,0)[r]{0.1}}
\put(890.0,335.0){\rule[-0.200pt]{4.818pt}{0.400pt}}
\put(221.0,451.0){\rule[-0.200pt]{4.818pt}{0.400pt}}
\put(199,451){\makebox(0,0)[r]{1}}
\put(890.0,451.0){\rule[-0.200pt]{4.818pt}{0.400pt}}
\put(221.0,568.0){\rule[-0.200pt]{4.818pt}{0.400pt}}
\put(199,568){\makebox(0,0)[r]{10}}
\put(890.0,568.0){\rule[-0.200pt]{4.818pt}{0.400pt}}
\put(221.0,684.0){\rule[-0.200pt]{4.818pt}{0.400pt}}
\put(199,684){\makebox(0,0)[r]{100}}
\put(890.0,684.0){\rule[-0.200pt]{4.818pt}{0.400pt}}
\put(221.0,68.0){\rule[-0.200pt]{165.980pt}{0.400pt}}
\put(910.0,68.0){\rule[-0.200pt]{0.400pt}{167.907pt}}
\put(221.0,765.0){\rule[-0.200pt]{165.980pt}{0.400pt}}
\put(45,416){\makebox(0,0){$m_{q}$ (GeV)}}
\put(359,153){\makebox(0,0)[l]{$u$}}
\put(457,187){\makebox(0,0)[l]{$d$}}
\put(556,338){\makebox(0,0)[l]{$s$}}
\put(654,465){\makebox(0,0)[l]{$c$}}
\put(753,524){\makebox(0,0)[l]{$b$}}
\put(851,709){\makebox(0,0)[l]{$t$}}
\put(221.0,68.0){\rule[-0.200pt]{0.400pt}{167.907pt}}
\put(285,123){\usebox{\plotpoint}}
\put(285.0,123.0){\rule[-0.200pt]{16.622pt}{0.400pt}}
\put(354.0,123.0){\rule[-0.200pt]{0.400pt}{14.695pt}}
\put(285.0,184.0){\rule[-0.200pt]{16.622pt}{0.400pt}}
\put(285.0,123.0){\rule[-0.200pt]{0.400pt}{14.695pt}}
\put(383,158){\usebox{\plotpoint}}
\put(383.0,158.0){\rule[-0.200pt]{16.622pt}{0.400pt}}
\put(452.0,158.0){\rule[-0.200pt]{0.400pt}{13.490pt}}
\put(383.0,214.0){\rule[-0.200pt]{16.622pt}{0.400pt}}
\put(383.0,158.0){\rule[-0.200pt]{0.400pt}{13.490pt}}
\put(481,310){\usebox{\plotpoint}}
\put(481.0,310.0){\rule[-0.200pt]{16.863pt}{0.400pt}}
\put(551.0,310.0){\rule[-0.200pt]{0.400pt}{12.527pt}}
\put(482.0,362.0){\rule[-0.200pt]{16.622pt}{0.400pt}}
\put(482.0,310.0){\rule[-0.200pt]{0.400pt}{12.527pt}}
\put(580,456){\usebox{\plotpoint}}
\put(580.0,456.0){\rule[-0.200pt]{16.622pt}{0.400pt}}
\put(649.0,456.0){\rule[-0.200pt]{0.400pt}{2.891pt}}
\put(580.0,468.0){\rule[-0.200pt]{16.622pt}{0.400pt}}
\put(580.0,456.0){\rule[-0.200pt]{0.400pt}{2.891pt}}
\put(679,523){\usebox{\plotpoint}}
\put(679.0,523.0){\rule[-0.200pt]{16.622pt}{0.400pt}}
\put(748.0,523.0){\rule[-0.200pt]{0.400pt}{0.723pt}}
\put(679.0,526.0){\rule[-0.200pt]{16.622pt}{0.400pt}}
\put(679.0,523.0){\rule[-0.200pt]{0.400pt}{0.723pt}}
\put(777,708){\usebox{\plotpoint}}
\put(777.0,708.0){\rule[-0.200pt]{16.622pt}{0.400pt}}
\put(846.0,708.0){\rule[-0.200pt]{0.400pt}{0.723pt}}
\put(777.0,711.0){\rule[-0.200pt]{16.622pt}{0.400pt}}
\put(777.0,708.0){\rule[-0.200pt]{0.400pt}{0.723pt}}
\end{picture}
\end{center}
\caption{The quark mass spectrum.  The bands indicate the  
running $\overline{\rm MS}$ mass, evaluated at the quark mass 
(for $c,b,t$) or 
at 2 GeV (for $u,d,s$), and the associated uncertainty.}  
\label{fig:massspectrum}
\end{figure}
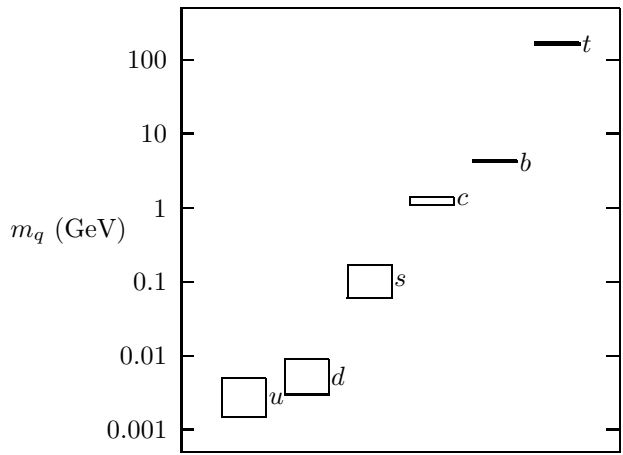

\section{Top mass}

The top-quark mass has been measured by the CDF~\cite{CDFMASS} and 
D0~\cite{D0MASS} collaborations to be
\begin{eqnarray}
m_t & = & 176.0 \pm 6.5\;{\rm GeV}\;({\rm CDF}) \\ 
    & = & 172.1 \pm 7.1\;{\rm GeV}\;({\rm D0})\;. 
\end{eqnarray}
This yields a world-average mass of~\cite{AVGMASS}\footnote{This is the 
top-quark pole mass. The corresponding $\overline{\rm MS}$ mass is 
$m_t^{\overline {\rm MS}}(m_t^{\overline {\rm MS}})
= 165.2 \pm 5.1$ GeV.\cite{GBGS}}
\begin{equation}
m_t = 174.3 \pm 5.1\;{\rm GeV}\;({\rm CDF} + {\rm D0})\;.
\end{equation}
To put this into context, I plot all the quark masses in Fig.~1, 
on a logarithmic scale. The width
of each band is proportional to the fractional uncertainty in the quark mass.  
We see that, at present, the top-quark mass is the best-known quark mass,
with the $b$-quark mass a close second 
($m_b^{\overline {\rm MS}}(m_b)
=4.25 \pm 0.15$ GeV).\cite{PDG}

\begin{figure}[t]
\begin{center}
\epsfxsize= 3.0in
\leavevmode
\epsfbox{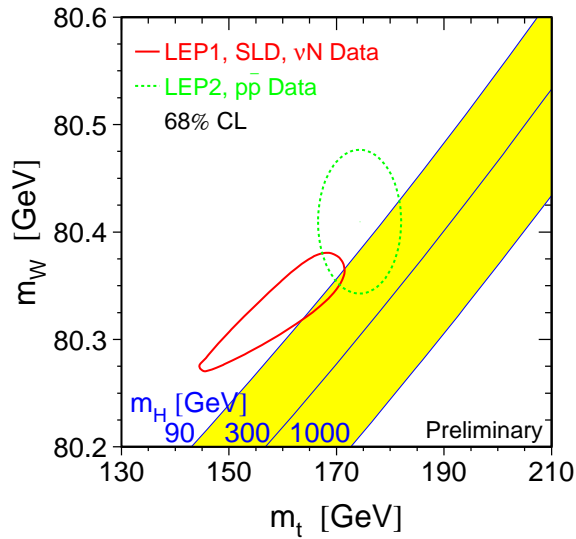}
\end{center}
\caption[fake]{$W$ mass {\it vs.}~top-quark mass, with lines of 
constant Higgs mass.
The solid ellipse is the $1\sigma$ contour from precision electroweak 
experiments.  The dashed ellipse is the $1\sigma$ contour from direct
measurements.  Figure from Ref.~\cite{LEPEWWG}.}
\end{figure}

An important question for the future is what precision we desire for the 
top-quark mass.  There are at least two avenues along which to address this
question.  One is in the context of precision electroweak data.
Fig.~2 summarizes the world's precision electroweak data on a plot of 
$M_W$ {\it vs.}~$m_t$.  
The solid ellipse is the $1\sigma$ contour.  If the
standard electroweak model is correct, the measured top-quark mass should
lie within this contour.  Since the contour spans about $\pm 8$ GeV 
along the $m_t$ axis, we conclude that the present uncertainty of $\pm 5$ GeV
in the top-quark mass is more than sufficient for the purpose of precision
electroweak physics at this time.  

There is one electroweak
measurement, $M_W$, whose precision could increase significantly.
An uncertainty of $\pm 20$ MeV is a realistic goal for Run III
at the Tevatron.\cite{TEV2000}
Let us take this uncertainty and project it onto
a line of constant Higgs mass in Fig.~2.  This is appropriate, 
because once a Higgs boson is discovered, even a crude knowledge of
its mass will define a narrow line in Fig.~2, since precision electroweak
measurements are sensitive only to the logarithm of the Higgs mass.  
An uncertainty in $M_W$ of $\pm 20$ GeV projected onto a line of constant
Higgs mass corresponds to an uncertainty of $\pm 3$ GeV in the top-quark
mass.  Thus we desire a measurement of $m_t$ to $\pm 3$ GeV 
in order to make maximal use of the precision measurement of $M_W$.

\begin{figure}[t]
\begin{center}
\setlength{\unitlength}{1.0in}
\begin{picture}(3.0,1.5)(0.0,0.0)
\epsfxsize= 1.5in
\leavevmode
\put(0.0,0.0){\epsfbox{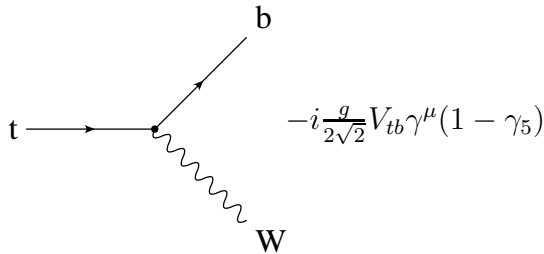}}
\put(1.5,0.88){{\large $-i \frac{g}{2 \sqrt{2}} V_{tb} \gamma^{\mu} 
(1-\gamma_{5})$}}
\end{picture}
\caption{Top-quark charged-current weak interaction.}
\end{center}
\end{figure}

Another avenue along which to address the desired accuracy of the top-quark 
mass is to
recall that the top-quark mass is a fundamental parameter of the standard
model.  Actually, it is the top-quark Yukawa coupling which is the 
fundamental parameter, given by
\begin{equation}
y_t = \sqrt 2 \frac{m_t}{v} \approx 1\;.
\label{yukawa}
\end{equation}
The fact that this coupling is of order unity suggests that it may be a
truly fundamental parameter.  We hope someday to have a theory that relates
the top-quark Yukawa coupling to that of its SU(2) partner, 
the $b$ quark.\footnote{A particularly compelling model which relates
the $b$ and top-quark masses is SO(10) grand 
unification.\cite{G} This model may be able to account for the 
masses of all the fourth-generation fermions, including the tau neutrino,
whose mass is given by the ``see-saw'' mechanism \cite{GRS} as
$m_{\nu_\tau} \approx m_t^2/M_{GUT} \approx 10^{-2}$ eV.\cite{W}} 
The $b$-quark mass is currently known with an accuracy of $\pm 3.5\%$.
Since the uncertainty is entirely theoretical, it is likely that it will
be reduced in the future.  If we assume that future work cuts the uncertainty 
in half, the corresponding uncertainty in the top-quark mass would be
$\pm 3$ GeV.

We conclude that both precision electroweak experiments and $m_t$ as a 
fundamental parameter lead us to the desire to measure the top-quark
mass with an accuracy of $\pm 3$ GeV.  This is well matched with 
future expectations.  An uncertainty of $\pm 3$ GeV per experiment 
is anticipated in Run II,\cite{RUNII} and additional running could
reduce this uncertainty to $\pm 2$ GeV.\cite{TEV2000}

\section{Top weak interaction}

The standard model dictates that the top quark has the same $V-A$ 
charged-current weak interaction as all the other fermions, as
shown in Fig.~3.
It is easy to see that this implies that the $W$ boson in top decay 
cannot be right handed, {\it i.e.}, positive helicity.   The argument is 
sketched in
Fig.~4.  In the limit of a massless $b$ quark, the $V-A$ current dictates
that the $b$ quark in top decay is always left-handed.\footnote{Being far
from massless, the decaying top quark can be left- or right-handed.}
If the $W$ boson were right-handed, then the component of total angular
momentum along the decay axis would be $+3/2$ (there is no component
of orbital angular momentum along this axis).  But the initial top quark
has spin angular momentum $\pm 1/2$ along this axis, so this decay is
forbidden by conservation of angular momentum.  
CDF has measured
\begin{equation}
BR(t\to W_+b) = 0.11 \pm 0.15 \pm 0.06
\end{equation}
which is consistent with zero.\cite{ROSER}

\begin{figure}[t]
\begin{center}
\epsfxsize= 2.1in
\leavevmode
\epsfbox{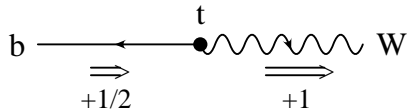}
\end{center}
\caption{Illustration that the top quark cannot decay to a right-handed
(positive-helicity) $W$ boson.}
\end{figure}

The top quark may decay to a left-handed (negative helicity) or a longitudinal
(zero helicity) $W$ boson.
Its coupling to a longitudinal $W$ boson is similar to its Yukawa coupling,
Eq.~(\ref{yukawa}), which is enhanced with respect to the weak coupling.
Therefore the top quark prefers to decay to 
a longitudinal $W$ boson, with a branching ratio
\begin{equation}
BR(t\to W_0 b) = \frac{m_t^2}{m_t^2+2M_W^2} \approx 70\%\;.
\end{equation}
CDF has made a first measurement of this branching ratio,\cite{WLONG} 
\begin{equation}
BR(t\to W_0 b) = 0.55 \pm 0.32 \pm 0.12\;,
\end{equation}
which is consistent with expectations.
The anticipated accuracy of this measurement in Run II and beyond will 
make it an interesting quantitative test of the top-quark weak 
interaction.\cite{TEV2000,RUNII}

\begin{figure}[t]
\begin{center}
\epsfxsize= 3.8in
\leavevmode
\epsfbox{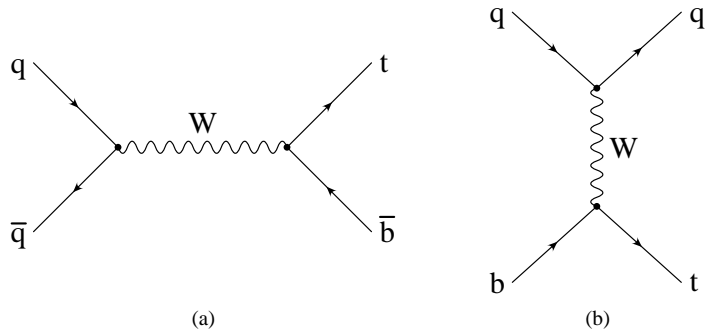}
\end{center}
\caption{Single-top-quark production via the weak interaction: 
(a) $s$-channel process; (b) $t$-channel process.}
\end{figure}

In addition to the $V-A$ structure of the top weak interaction, there is also 
its strength, {\it i.e.}, the value of the Cabibbo-Kobayashi-Maskawa
(CKM) matrix element $V_{tb}$.  CDF has measured \cite{WLONG}
\begin{equation}
\frac{BR(t\to Wb)}{BR(t\to Wq)} 
= \frac{|V_{tb}|^2}{|V_{td}|^2+|V_{ts}|^2+|V_{tb}|^2} = 0.99 \pm 0.29
\label{ratio}
\end{equation}
and it is interesting to ask what this tells us about
$V_{tb}$.  If we assume that there are just three generations of quarks,
then unitarity of the CKM matrix implies that the denominator of 
Eq.~(\ref{ratio}) is unity, and we can immediately extract
\begin{equation}
|V_{tb}| = 0.99 \pm 0.15 \;(> 0.76\;{\rm 95\%\;CL}) \;(3\;{\rm generations}).
\end{equation}
However, to put this into perspective, recall that three-generation 
unitarity also implies that 
$|V_{ub}|^2+|V_{cb}|^2+|V_{tb}|^2=1$, and since $|V_{ub}|$ and $|V_{cb}|$
have been measured to be small, one finds \cite{PDG}
\begin{equation}
|V_{tb}| = 0.9991 - 0.9994 \;(3\;{\rm generations})
\end{equation}
which is far more accurate than the CDF result.

If we assume more than three generations, then unitarity implies almost
nothing about $|V_{tb}|$:~\cite{PDG}
\begin{equation}
|V_{tb}| = 0 - 0.9994\;(>3\;{\rm generations})\;.
\end{equation}
At the same time, we also lose the constraint that the denominator of
the middle expression in Eq.~(\ref{ratio}) is unity.  
All we can conclude from Eq.~(\ref{ratio})
is that $|V_{tb}| >> |V_{ts}|,|V_{td}|$; we learn nothing about its
absolute magnitude.  

Fortunately, there is a direct way to measure $|V_{tb}|$ at the Tevatron, 
which makes no assumptions about the number of generations.  One uses
the weak interaction to produce the top quark; the two relevant processes
are shown in Fig.~5.  The cross sections for these two ``single top''
processes are proportional to $|V_{tb}|^2$.  The first process
involves an $s$-channel $W$ boson,\cite{CP} while
the second process involves
a $t$-channel $W$ boson (and is often called $W$-gluon fusion, because
the initial $b$ quark actually comes from a gluon splitting 
to $b\bar b$).\cite{DW}
$W$-gluon fusion has the advantage of greater statistics than the 
$s$-channel process,
but the disadvantage of greater theoretical uncertainty.  
Thus far there is only a bound on single-top-quark production via
$W$-gluon fusion from CDF,\cite{WLONG} 
\begin{equation}
\sigma(Wg\to t\bar b) < 15.4\;{\rm pb}
\end{equation}
which is an order of magnitude away from the theoretical expectation 
of $1.70 \pm 0.09$ pb.\cite{SSWNLO}  Both single-top processes should 
be observed in Run II.\cite{TEV2000,RUNII}

\begin{figure}[t]
\begin{center}
\epsfxsize= 4.0in
\leavevmode
\epsfbox{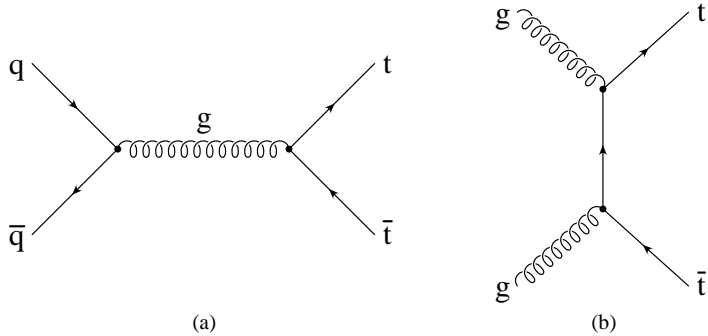}
\end{center}
\caption{Top-quark pair production via the strong interaction:
(a) quark-antiquark annihilation; (b) gluon fusion.}
\end{figure}

Single-top-quark production can also be used to test the $V-A$ structure
of the top-quark charged-current weak interaction.  This structure implies that
the top-quark spin is 100\% polarized along the direction (in the 
top-quark rest frame) of the $d$ or $\bar d$ quark in the event, 
in both $W$-gluon fusion and the $s$-channel process.\cite{MP2}

\section{Top strong interaction}

The strong interaction of the top quark is best tested in its production.
There are two subprocesses by which $t\bar t$ pairs are produced by the 
strong interaction at a 
hadron collider, shown in Fig.~6.  At the Tevatron, the quark-antiquark
annihilation process is dominant, accounting for $90\%$ of the cross 
section at $\sqrt S = 1.8$ TeV.  When the machine energy is increased
to $\sqrt S = 2$ TeV in Run II, this fraction decreases to $85\%$.  
The cross section increases considerably, by about $37\%$, when the
machine energy is increased from 1.8 to 2 TeV.

\begin{figure}[t]
\begin{center}
\epsfysize=6.5 cm
\leavevmode
\epsfbox[35 150 530 655]{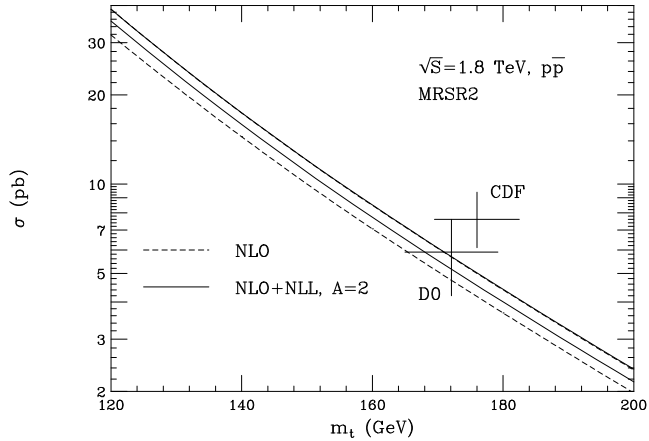} 
\end{center}
\caption[fake]{Cross section for $t\bar t$ production at the Tevatron 
{\it vs.}~the top-quark mass.  Dashed band is from next-to-leading-order QCD;
solid band includes soft-gluon resummation at next-to-leading-logarithm.
Figure adapted from Ref.~\cite{BCMN}.}
\label{fig:topcross}
\end{figure}

We show in Fig.~7 the $t\bar t$ cross section {\it vs.}~the top-quark mass.
The dashed band is from a calculation at next-to-leading-order in QCD.
\cite{NDE,BCMN} The uncertainty in this calculation is about $\pm 10\%$.
The solid band includes the effect of soft gluon resummation at 
next-to-leading logarithm; this increases the cross section by only a few
percent, but reduces the uncertainty by almost 
a factor of 2.\cite{BCMN}\footnote{These bands reflect the uncertainty in
the cross section 
due to the variation of the renormalization and factorization scales.
They do not include the uncertainty
from $\alpha_s(M_Z)$ or the parton distribution functions.  However, these
additional uncertainties are relatively modest.\cite{CMNT}}    
The measurements by CDF and D0,
\begin{eqnarray}
\sigma & = & 7.6^{+1.8}_{-1.5}\; {\rm pb}\;({\rm CDF}) \\
\sigma & = & 5.9 \pm 1.7\; {\rm pb}\;({\rm D0})
\end{eqnarray}
are also shown in the figure, and are seen to agree with theory within 
one standard deviation.  

\begin{figure}[t]
\begin{center}
\epsfxsize= 4.0in
\leavevmode
\epsfbox{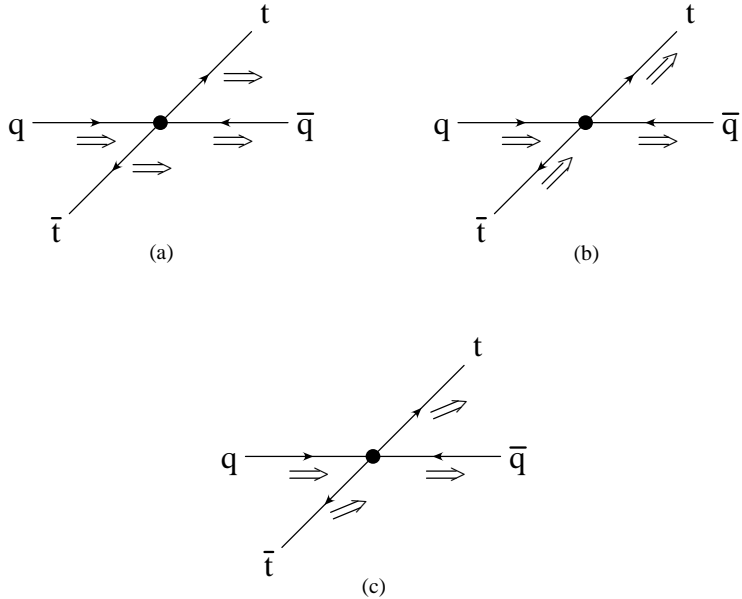}
\end{center}
\caption{Top-quark and light-quark spins in $q\bar q \to t\bar t$:
(a) near threshold; (b) far above threshold; (c) intermediate energies.}
\end{figure}

An interesting aspect of the strong production of $t\bar t$ pairs is that
the spins of the $t$ and $\bar t$ are nearly $100\%$ correlated.\cite{MP}
The correct basis in which to measure the spins requires some consideration,
however.  At threshold ($\sqrt s \approx 2m_t$), the cross section is
entirely $s$ wave, so the spins of the colliding quarks are transferred
to the $t$ and $\bar t$.  Since the quark-antiquark
annihilation takes place via a gauge interaction, the quark and antiquark
must have opposite helicities, so the spins of the $t$ and $\bar t$ 
are aligned along the beamline as shown in Fig.~8(a).  At the other
extreme, far above threshold ($\sqrt s >> 2m_t$), the $t$ and $\bar t$ 
behave like massless quarks, and therefore must have opposite helicities,
as shown in Fig.~8(b).
The question is whether there is a basis which interpolates between the 
beamline basis near threshold and the helicity basis far above threshold,
and the answer is affirmative - it has been dubbed the ``off-diagonal'' 
basis.\cite{PS}  The $t$ and $\bar t$ spins are $100\%$ correlated in this
basis, as shown in Fig.~8(c). 
Since the quark-antiquark annihilation process accounts for most
of the cross section at the Tevatron, the spin correlation is nearly $100\%$.
This effect should be observable in Run II.

Another interesting aspect of the strong production of $t\bar t$ pairs is 
an asymmetry in the distribution of the $t$ and $\bar t$ quarks.\cite{KR}  
This effect arises at one loop, and leads to a forward-backward asymmetry 
of about 5\% in $t\bar t$ production at the Tevatron.

\section{Rare decays}

\begin{figure}[t]
\begin{center}
\epsfxsize= 3.40in
\leavevmode
\epsfbox{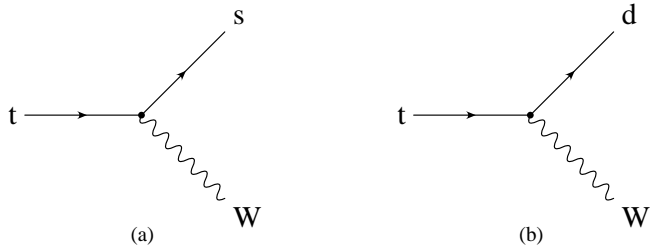}
\end{center}
\caption{Rare top decays: (a) $t\to Ws$; (b) $t\to Wd$.}
\end{figure}

Rare top decays in the standard model tend to be very rare, outside the
range of the Tevatron.  Thus far CDF has placed limits on the rare decays
\cite{TZC}
\begin{eqnarray}
BR(t\to Zq) & < & 33\%\;(95\% {\rm CL})\\
BR(t\to \gamma q) & < & 3.2\%\;(95\% {\rm CL})
\end{eqnarray}
which have tiny branching ratios in the standard model.\cite{EHS}

The least rare of the rare decays within the standard model are the CKM 
suppressed decays $t\to Ws$ and $t\to Wd$, shown in Fig.~9.  These decays
are interesting because they allow a direct measurement of the CKM matrix
elements $V_{ts}$ and $V_{td}$.  Assuming three generations, the branching 
ratios are predicted to be
\begin{eqnarray}
BR(t\to Ws) & \approx & 0.1\% \\
BR(t\to Wd) & \approx & 0.01\%
\end{eqnarray}
which are small, but not tiny.  Since there will be about 15,000 raw $t\bar t$
pairs produced in Run II, and about 200,000 in Run III, events of these
type will be present in the data.  However, there is no 
generally-accepted strategy for identifying these events.

\section*{Acknowledgments}

I am grateful for conversations with and assistance from M.~Mangano,
K.~Paul, R.~Roser, S.~Snyder, and T.~Stelzer.
This work was supported in part by Department of Energy grant 
DE-FG02-91ER40677.

\end{document}